\documentclass[conference,letterpaper]{IEEEtran}

\usepackage[utf8]{inputenc} 
\usepackage[T1]{fontenc}
\usepackage{url}
\usepackage{ifthen}
\usepackage{cite}
\usepackage[cmex10]{amsmath} 

\usepackage{booktabs}       
\usepackage{amsfonts}       
\usepackage{nicefrac}       
\usepackage{microtype}      
\usepackage{xcolor}         
\usepackage{amsmath}
\usepackage{amsthm}
\usepackage{graphicx}
\usepackage{caption}
\usepackage{balance}

\newtheorem{theorem}{Theorem}
\newtheorem{lemma}{Lemma}
\newtheorem{proposition}{Proposition}

\newtheorem{definition}{Definition}

\usepackage{url}

\newcommand*{\Scale}[2][4]{\scalebox{#1}{$#2$}}%

\begin{document}
\title{Distributed Learning for Dynamic Congestion Games} 

\author{%
 \IEEEauthorblockN{Hongbo Li, and Lingjie Duan}
 \IEEEauthorblockA{Singapore University of Technology and Design, Singapore\\
                   Email: hongbo\_li@mymail.sutd.edu.sg; lingjie\_duan@sutd.edu.sg}
}



\maketitle


\begin{abstract}
Today mobile users learn and share their traffic observations via crowdsourcing platforms (e.g., Google Maps and Waze). Yet such platforms myopically recommend the currently shortest path to users, and selfish users are unwilling to travel to longer paths of varying traffic conditions to explore. Prior studies focus on one-shot congestion games without information learning, while our work studies how users learn and alter traffic conditions on stochastic paths in a distributed manner. Our analysis shows that, as compared to the social optimum in minimizing the long-term social cost via optimal exploration-exploitation tradeoff, the myopic routing policy leads to severe under-exploration of stochastic paths with the price of anarchy (PoA) greater than \(2\). Besides, it fails to ensure the correct learning convergence about users' traffic hazard beliefs. To mitigate the efficiency loss, we first show that existing information-hiding mechanisms and deterministic path-recommendation mechanisms in Bayesian persuasion literature do not work with even \(\text{PoA}=\infty\). Accordingly, we propose a new combined hiding and probabilistic recommendation (CHAR) mechanism to hide all information from a selected user group and provide state-dependent probabilistic recommendations to the other user group. Our CHAR successfully ensures PoA less than \(\frac{5}{4}\), which cannot be further reduced by any other informational mechanism. Additionally, we experiment with real-world data to verify our CHAR's good average performance.
\end{abstract}

\section{Introduction}
\label{Section1}
In today's traffic networks, users sequentially arrive and choose their routing paths based on real-time traffic conditions \cite{zhang2022review}. To help new user arrivals, crowdsourcing platforms like Google Maps and Waze aggregate the latest traffic information from former users who just traveled there and shared their observations \cite{waze}. However, these platforms myopically recommend only the shortest path, and selfish users are not willing to explore longer paths of varying traffic conditions. Prior congestion game literature assumes a static routing scenario, overlooking users' long-term information learning (\!\! \cite{zhu2022information,vasserman2015implementing}).

To efficiently learn and share information, multi-armed bandit (MAB) problems are developed to study the optimal exploration-exploitation among stochastic arms/paths (e.g., \cite{liu2012learning,slivkins2019introduction,gupta2021multi}). Recent studies extend to distributed MAB scenarios with multiple user arrivals at the same time (e.g., \cite{shi2021federated,yang2021cooperative,zhu2023distributed}). For instance, \cite{shi2021federated} explores cooperative information exchange among all users to facilitate local model learning.
Subsequently, both \cite{yang2021cooperative} and \cite{zhu2023distributed} examine partially connected communication graphs, where users communicate only with neighbors to make locally optimal decisions. In general, these MAB solutions assume users' cooperation to align with the social planner, without possible selfish deviations to improve their own benefits.

As selfish users may not listen to the social planner, recent works focus on information design like Bayesian persuasion to incentivize users' exploration \cite{mansour2022bayesian,babichenko2022regret,gollapudi2023online}. For example, \cite{mansour2022bayesian} properly discloses the collected information to regulate user arrivals and control efficiency loss. Without perfect feedback, \cite{babichenko2022regret} considers all possible user utilities of traveling each path to improve the mechanism's robustness. However, all these works assume that users' routing decisions do not internally alter the traffic congestion condition for future users, and only focus on exogenous information to learn dynamically.  

It is practical to consider endogenous information variation in dynamic congestion games, where more users choosing a path not only improve learning accuracy there (\!\!\cite{yang2009discretization}), but also produce more congestion (\!\!\cite{meunier2010equilibrium,carmona2020pure}) for followers. For this new problem, we need to overcome two technical issues. The first is how to \emph{dynamically allocate users to reach the best trade-off between learning accuracy and resultant congestion costs of stochastic paths}. To minimize the long-run social travel cost, it is critical to dynamically balance learning and congestion effects to avoid under- or over-exploration of stochastic paths. 

Even provided with the socially optimal policy, we still need to ensure selfish users follow it. Thus, the second issue is how to \emph{design an informational mechanism to properly regulate users' myopic routing}. In practice, informational mechanisms (e.g., Bayesian persuasion \cite{das2017reducing,kamenica2019bayesian,mansour2022bayesian}) are non-monetary and easier to implement compared to pricing (e.g., \cite{li2022online,Ferguson2022effective}). Two commonly informational mechanisms used to optimize long-run information learning in congestion games are information hiding \cite{tavafoghi2017informational,wang2020efficient,farhadi2022dynamic} and deterministic path-recommendation \cite{li2019recommending,wu2019learning,li2023congestion}. We will show later that both mechanisms do not work in improving our system performance, inspiring our new design.

Our main contributions are summarized as follows. 
\begin{itemize}
    \item \emph{Novel distributed learning for dynamic congestion games:} In Section \ref{section2}, we study users' learning and routing to alter traffic conditions on stochastic paths. In this new problem, more users' routing on stochastic paths generates both positive learning benefits and negative congestion effects for followers. We use users’ positive learning to negate negative congestion, which fundamentally extends traditional one-shot congestion games (e.g., \cite{mansour2022bayesian,gollapudi2023online}).
    \item \emph{Policies comparison via PoA analysis:} 
    To minimize the social cost, in Section \ref{section3}, we formulate optimization problems for both myopic and socially optimal policies as a Markov decision process (MDP). Compared to the social optimum in minimizing the long-term social cost via optimal exploration-exploitation tradeoff, the myopic routing policy (used by Google Maps and Waze) leads to severe under-exploration of stochastic paths with the price of anarchy (PoA) greater than \(2\).
    \item \emph{Learning convergence and new mechanism design:} In Section \ref{section5}, we first prove that the socially optimal policy ensures correct convergence of users’ traffic hazard beliefs on stochastic paths, while the myopic policy cannot. Then we show that existing information-hiding and deterministic-recommendation mechanisms in Bayesian persuasion literature make $\Scale[0.94]{\text{PoA}=\infty}$. Accordingly, we propose a combined hiding and probabilistic recommendation (CHAR) mechanism to hide all information from a selected user group and provide state-dependent probabilistic recommendations to the other user group, achieving the minimum possible $\text{PoA}<\frac{5}{4}$.
\end{itemize}

\section{System Model}\label{section2}
\begin{figure}[t]
    \centering
    \includegraphics[width=0.28\textwidth]{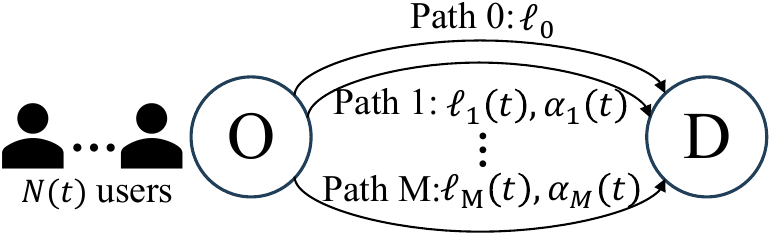}
    \captionsetup{font={footnotesize}}
    \caption{Dynamic congestion model: within each time slot $t$, a random number $\Scale[0.92]{N(t)}$ of users arrive at O to decide routings to D.}
    \label{fig:congestion_game}
\end{figure}

In this section, we fist introduce the dynamic congestion model for a typical parallel multi-path network as in existing congestion game literature (e.g., \cite{tavafoghi2017informational, wu2019learning} and \cite{li2023congestion}). Then we introduce the distributed learning model for the crowdsourcing platform.
As shown in Fig.~\ref{fig:congestion_game}, we consider an infinite discrete time horizon $\Scale[0.94]{t\in\{1,2,\cdots\}}$ to model the dynamic congestion game. Within each time slot $t$, a stochastic number $N(t)\in\ \Scale[0.92]{\{N_{min},\cdots,N_{max}\}}$ of atomic users, where $\Scale[0.92]{\mathbb{E}[N(t)]=N}$, arrive at origin O to decide their paths to travel to destination D.

\subsection{Dynamic Congestion Model}

Following the traditional routing game literature (\!\!\cite{kremer2014implementing,das2017reducing,li2023congestion}), we model a safe path 0 with fixed travel latency $\ell_0$, and let $\ell_i(t)$ denote the varying travel latency of stochastic path $i$ at the beginning of time $t\in\{1,2,\cdots\}$. According to \cite{ban2009delay} and \cite{alam2019prediction}, $\ell_i(t+1)$ at $t+1$ is correlated with both current $\ell_i(t)$ and the number of atomic users traveling this path, which is denoted by $n_i(t)$. For example, more users traveling on the same path results in slower driving speeds, thereby increasing the latency for later users. Define this general correlation function as:
\begin{align}\label{ell_2}
    \ell_i(t+1)=f\big(\ell_i(t),n_i(t),\alpha_i(t)\big),
\end{align}
where $\alpha_i(t)$ is the correlation coefficient, measuring the leftover flow to be served over time. Note that correlation function $f(\cdot)$ is a general increasing function in $\ell_i(t),n_i(t)$ and $\alpha_i(t)$.

As in existing congestion game literature (\!\!\cite{wu2019learning,farhadi2022dynamic,li2023congestion}), we model $\alpha_i(t)$ to follow a memoryless stochastic process (not necessarily a time-invariant Markov chain) to alter between a high hazard (i.e., bad) state $\Scale[0.94]{\alpha_H\in[1,+\infty)}$ and a low hazard (good) state $\Scale[0.94]{\alpha_L\in [0,1)}$.
Unlike these works assuming that the social planner knows the probability distribution of $\alpha_i(t)$ beforehand, we allow it to be unknown in our model here. 

Denote the expected steady state of $\alpha_i(t)$ in the long run by
\begin{align}
    \mathbb{E}[\alpha_i(t)|\Bar{x}]=\Bar{x}\cdot \alpha_H+(1-\Bar{x})\cdot \alpha_L, \label{bar_alpha}
\end{align}
where $\Bar{x}$ is the long-run expected probability of $\alpha_i(t)=\alpha_H$ yet the exact value of $\Bar{x}$ is unknown to users. They can only know $\Bar{x}$ to satisfy a distribution $\mathbb{P}(\Bar{x})$. To accurately estimate $\mathbb{E}[\alpha_i(t)|\Bar{x}]$, the system expects users to learn the real steady state $\Bar{x}$ by observing real-time traffic conditions.

\subsection{Distributed Learning Model}
When traveling on stochastic path $i$, users cannot observe $\alpha_i(t)$ directly but a traffic hazard event (e.g., jamming). At the same time, \cite{venanzi2013crowdsourcing} shows that each user has a noisy observation of the hazard. Hence, our system uses a majority vote to fuse all their observation reports of stochastic path $i$ into a hazard summary $\Scale[0.94]{y_i(t)\in\{0,1,\emptyset\}}$ during time $t$. Specifically, $\Scale[0.95]{y_i(t) = 1}$ tells that most of the current $\Scale[0.95]{n_i(t)}$ users observe a traffic hazard on stochastic path $i$. $\Scale[0.95]{y_i(t)=0}$ tells that most users observe no hazard on path $i$. $\Scale[0.95]{y_i(t)=\emptyset}$ tells no user observation on path~$i$. 

However, the fused summary $y_i(t)$ by $n_i(t)$ users can still be inaccurate. Given $n_i(t)$ and high-hazard $\alpha_i(t)=\alpha_H$, we define the group probability for observing a hazard ($y_i(t)=1$) to be $\Scale[0.94]{q_H(n_i(t))=\mathbf{Pr}\big(y_i(t)=1|\alpha_i(t)=\alpha_H, n_i(t)\big).}$
Similarly, let $\Scale[0.93]{q_L(n_i(t))\in[0,1]}$ denote the corresponding probability under low-hazard $\alpha_L$.
According to \cite{yang2009discretization}, $\Scale[0.94]{q_H(n_i(t))\in[0,1]}$ increases with $n_i(t)$, because more users help spot hazard $\Scale[0.94]{\alpha_i(t)=\alpha_H}$. Similarly, $\Scale[0.94]{q_L(n_i(t))}$ decreases with $n_i(t)$ under $\Scale[0.94]{\alpha_i(t)=\alpha_L}$. 

Following the memoryless property \cite{li2019recommending}, we use Bayesian inference to equivalently summarize users' routing decisions and observations into a prior hazard belief $\Scale[0.94]{x_i(t)\in[0,1]}$ for any stochastic path~$i$, indicating the probability of $\Scale[0.94]{\alpha_i(t)=\alpha_H}$ at $t$:
\begin{align}
    \Scale[0.94]{x_i(t)=\mathbf{Pr}\big(\alpha_i(t)=\alpha_H|x_i(t-1),y_i(t-1),n_i(t-1)\big)}.\label{x(t)}
\end{align}


At the beginning of time $t$, the platform publishes latest $\mathbb{E}[\ell_i(t)|x_i(t-1),y_i(t-1)]$ and $x_i(t)$ on any stochastic path $i$. 

During time $t$, $\Scale[0.94]{N(t)}$ users arrive to decide $n_i(t)$ on each stochastic path~$i$ and travel there to return their observation summary $y_i(t)$. Based on $y_i(t)$, the platform next updates prior belief $x_i(t)$ to a posterior belief $x_i'(t)=\mathbf{Pr}(\alpha_i(t)=\alpha_H|x_i(t),y_i(t),n_i(t))$. For example, if $\Scale[0.94]{y_i(t)=1}$, we have
    \begin{align}       
        \Scale[0.94]{x_i'(t)}=\frac{\Scale[0.94]{x_i(t)q_H(n_i(t))}}{\Scale[0.94]{x_i(t)q_H(n_i(t))+(1-x_i(t))q_L(n_i(t))}}\label{x'(t)}
    \end{align}
    by Bayes' Theorem. If $y_i(t)=0$, we similarly calculate $x_i'(t)$, and we keep $\Scale[0.94]{x_i'(t)=x_i(t)}$ if $\Scale[0.94]{y_i(t)=\emptyset}$ without observation.

At the end of time $t$, the platform updates the expected correlation coefficient given posterior belief $x'_i(t)$:
    \begin{align}
        \Scale[0.94]{\mathbb{E}[\alpha_i(t)|x_i'(t)]=x_i'(t)\alpha_H+(1-x_i'(t))\alpha_L.}\label{E_alpha}
    \end{align}
    Combing (\ref{E_alpha}) with (\ref{ell_2}), we obtain expected travel latency
    \begin{align}
        \Scale[0.92]{
        \mathbb{E}[\ell_i(t+1)|x_i'(t)]=\mathbb{E}\big[f\big(\mathbb{E}[\ell_i(t)|x'_i(t-1)],n_i(t),\alpha_i(t)\big)\big]}\label{E_ell_2}
    \end{align}
     on path 2 for next time slot $\Scale[0.94]{t+1}$.
    Finally, the platform updates $\Scale[0.94]{x_i(t+1)=x_i'(t)}$ and repeats  above process in $t + 1$.

\section{Problem Formulations and Policies Comparison}\label{section3}

In this section, we first formulate the optimization problems for both myopic and socially optimal policies. Then we analyze and compare these two policies via PoA analysis.

\subsection{Problem Formulation for Myopic Policy}
In this subsection, we focus on the myopic policy used by Google Maps and Waze, under which users aim to minimize their own travel costs. First, we summarize stochastic paths' expected latencies and hazard beliefs into vectors $\mathbf{L}(t)=\{\mathbb{E}[\ell_i(t)|x_i'(t-1)]|i\in\{1,\cdots,M\}\}$ and $\mathbf{x}(t)=\{x_i(t)|i\in\{1,\cdots,M\}\}$, respectively. Define $n_i^{(m)}(t)\in\{0,\cdots, N(t)\}$ to be the number of users on any path~$\Scale[0.94]{i\in\{0,\cdots, M\}}$ under myopic policy, and let vector $\Scale[0.94]{\mathbf{n}^{(m)}(t)}$ summarize all the $\Scale[0.94]{n_i^{(m)}(t)}$.

As in \cite{das2017reducing,tavafoghi2017informational} and \cite{wu2019learning}, for each of the $n_0(t)$ users on safe path 0, his expected travel cost consists of travel latency $\ell_0$ and congestion cost $\Scale[0.94]{n_0(t)}$ caused by others on this path:
\begin{align}
    \Scale[0.94]{c_0(n_0(t))=\ell_0+n_0(t).}\label{cm_1}
\end{align}
While for each user on stochastic path $i\in\{1,\cdots,N\}$, besides $\Scale[0.94]{\mathbb{E}[\ell_i(t)|x_i'(t-1)]}$ in (\ref{E_ell_2}) and $\Scale[0.94]{n_i(t)}$, he faces an extra error cost $\Scale[0.94]{V(n_i(t-1))}$ due to former $\Scale[0.94]{n_i(t-1)}$ users' imperfect observation summary $y_i(t-1)$ (\!\!\cite{smith2018bayesian}). $V(n_i(t-1))$ tells how much the gap (between the ${n_i(t-1)}$ reports fused result and the reality) adds to the total cost by misleading decision-making of future users. Then his travel cost is:
\begin{align}
    \Scale[0.93]{c_i(n_i(t))=\label{cm_2}\mathbb{E}[\ell_i(t)|x_i'(t-1)]+n_i(t)+V\big(n_i(t-1)\big),}
\end{align}
where $\Scale[0.92]{V(n_i(t-1))}$ is a general decreasing function of $\Scale[0.92]{n_i(t-1)}$, as more users improve learning accuracy on path $i$.

Based on individual costs (\ref{cm_1}) and (\ref{cm_2}), we next analyze myopic policy $n_i^{(m)}(t)$. For ease of exposition, we first consider $M=1$ in a two-path network to solve and explain $\Scale[0.94]{n_1^{(m)}(t)}$. For any path number $\Scale[0.94]{M\geq 2}$, we can similarly compute $\Scale[0.94]{n_i^{(m)}(t)}$ by balancing expected travel costs among all paths as in (\ref{nm(t)}).
\begin{lemma}
Under the myopic policy, given $\Scale[0.94]{\mathbb{E}[\ell_1(t)|x_1'(t-1)]}$ and $\Scale[0.94]{x_1(t)}$ of stochastic path 1, the exploration number is:
\begin{align}
    \Scale[0.94]{n^{(m)}_1(t)=}\begin{cases}
        \Scale[0.94]{0}, \quad\quad\quad\quad\ \ \quad\text{if }\Scale[0.94]{c_1(0)\geq c_0(N(t))},\\
        \Scale[0.94]{N(t)}, \quad\quad\quad\quad\text{if }\Scale[0.94]{c_1(N(t))\leq c_0(0)},\\
        \frac{\Scale[0.94]{N(t)}}{\Scale[0.94]{2}}+\frac{\Scale[0.94]{c_0(0)- c_1(0)}}{\Scale[0.94]{2}},\quad\text{ otherwise}.
    \end{cases}\label{nm(t)}
\end{align}
\end{lemma}
If path~1's minimum expected travel cost is larger than path 0's maximum cost, all users will choose path 0 with $\Scale[0.94]{n_1^{(m)}(t)=0}$ in the first case of (\ref{nm(t)}). Otherwise, in the last two cases of (\ref{nm(t)}), there is always a positive $\Scale[0.94]{n_1^{(m)}(t)\geq 1}$ number of users traveling on stochastic path 1 to learn and update information.

Based on the above analysis, we further examine the long-run social cost of all users since time $t$. For current $\Scale[0.94]{N(t)}$ user arrivals, their immediate social cost under the myopic policy is 
\begin{align}
    \Scale[0.94]{c(\mathbf{n}^{(m)}(t))=\sum_{i=0}^Mn_i^{(m)}(t)c_i(n_i^{(m)}(t))}\label{cm}
\end{align}
with $c_i(\cdot)$ in (\ref{cm_1}) or (\ref{cm_2}).
Define $\Scale[0.94]{C^{(m)}(\mathbf{L}(t),\mathbf{x}(t),N(t))}$ to be the long-term $\rho$-discounted social cost function. Based on (\ref{cm}), we leverage the Markov decision process (MDP) to formulate:
\begin{align}
    &\Scale[0.94]{C^{(m)}\big(\mathbf{L}(t),\mathbf{x}(t),N(t)\big)=}\label{Cm}\\ &\Scale[0.94]{c(\mathbf{n}^{(m)}(t))+\rho \mathbb{E}\big[C^{(m)}\big(\mathbf{L}(t+1),\mathbf{x}(t+1),N(t+1)\big)}\big],\notag
\end{align}
where the update of $\Scale[0.94]{\mathbb{E}[\ell_i(t+1)|x_i'(t)]}$ in (\ref{x'(t)}) and $x_i(t+1)$ in (\ref{E_ell_2}) depend on current $\Scale[0.94]{N(t)}$ users' routing decision $\Scale[0.94]{n_i^{(m)}(t)}$ and observation $\Scale[0.94]{y_i(t)}$ on path $i$. Here discount factor $\Scale[0.94]{\rho\in(0,1)}$ is widely used to discount future costs to present value \cite{roughgarden2005selfish}.

\subsection{Problem Formulation for Socially Optimal Policy}
Define $\Scale[0.94]{n_i^*(t)\in\{0,1,\cdots,N(t)\}}$ to be the exploration number under the socially optimal policy, and let vector $\mathbf{n}^*(t)$ summarize all the $n_i^*(t)$. Different from the myopic policy that only minimizes each user's travel cost, the social optimum aims to well control the overall congestion and information learning to minimize long-run expected social cost $\Scale[0.94]{C^*(\mathbf{L}(t),\mathbf{x}(t),N(t))}$.

Then we similarly leverage MDP to formulate the long-term objective function under the socially optimal policy as:
\begin{align}
    &\Scale[0.94]{C^*\big(\mathbf{L}(t),\mathbf{x}(t),N(t)\big)=}\label{C*}\\ &\min_{\mathbf{n}^*(t)}\ \ \Scale[0.94]{c(\mathbf{n}^*(t))+\rho \mathbb{E}\big[C^*\big(\mathbf{L}(t+1),\mathbf{x}(t+1),N(t+1)\big)\big]}, \notag
\end{align}
where the immediate social cost $c(\mathbf{n}^*(t))$ is similarly defined as in (\ref{cm}).
Note that (\ref{C*}) is non-convex and difficult to solve due to the curse of dimensionality under infinite time horizon \cite{bellman1966dynamic}. However, we still manage to derive some structural results to compare (\ref{C*}) to (\ref{Cm}) under the myopic policy in the following.

\subsection{Policies Comparison via PoA Analysis}
In this subsection, we prove that the myopic policy misses both exploration and exploitation over time, as compared to the social optimum, leading to $\text{PoA}\geq 2$. This implies at least doubled total travel cost and motivates our mechanism design in Section~\ref{section5}. Before that, we first analyze the monotonicity of the two policies $n_i^{(m)}(t)$ and $n_i^*(t)$ in the next lemma.
\begin{lemma}\label{lemma:n*-nm}
Both exploration numbers $\Scale[0.94]{n_i^{(m)}(t)}$ and $\Scale[0.94]{n_i^*(t)}$ decrease with $\Scale[0.94]{x_i(t)}$ and $\Scale[0.94]{V(\cdot)}$. While the difference $n_i^*(t)-n_i^{(m)}(t)$ increases with $\Scale[0.94]{x_i(t)}$ but decreases with $\Scale[0.94]{V(\cdot)}$.
\end{lemma}
Intuitively, 
as $\Scale[0.94]{V(\cdot)}$ enlarges, users' explorations will incur extra error costs, making both policies unwilling to explore.

Thanks to Lemma \ref{lemma:n*-nm}, we next prove that the myopic policy misses both proper exploration and exploitation of stochastic path $i$, as hazard belief dynamically changes over time.
\begin{proposition}\label{Prop:explore}
There exists a belief threshold $\Scale[0.94]{x_{th}\in(0,1)}$, such that the myopic policy will over-explore stochastic path $i$ (with $\Scale[0.94]{n_i^{(m)}(t)\geq n_i^*(t)}$) if $\Scale[0.94]{x_i(t)<x_{th}}$, and will under-explore (with $\Scale[0.94]{n_i^{(m)}(t)<  n_i^*(t)}$) if $\Scale[0.94]{x_i(t)\geq x_{th}}$, as compared to the socially optimal policy. This belief threshold $\Scale[0.94]{x_{th}}$ increases with $\Scale[0.94]{V(\cdot)}$.
\end{proposition}
If there is a strong hazard belief $\Scale[0.94]{x_i(t)\geq x_{th}}$ of $\Scale[0.94]{\alpha_i(t)=\alpha_H}$, the myopic policy is not willing to explore stochastic path $i$ due to longer latency. However, the social optimum still allocates some users to learn possible $\Scale[0.94]{\alpha_L}$ future others to exploit. In contrast, given weak $\Scale[0.94]{x_i(t)<x_{th}}$, myopic users flock to path~$i$ without considering future congestion, while the social optimum may exploit safe path 0 to reduce congestion on path~$i$. 

To examine the efficiency gap, we define the price of anarchy (PoA) as the maximum ratio between the social costs (\ref{Cm}) under the myopic policy and (\ref{C*}) under the social optimum \cite{roughgarden2005selfish}:
\begin{align}
    \text{PoA}^{(m)}=\max_{\begin{aligned}&\Scale[0.65]{\alpha_H,\alpha_L,\mathbf{x}(t),\ell_0,N(t)}\\[-6pt]&\Scale[0.65]{\ \ \mathbf{L}(t),q_H,q_L,V(\cdot)}\end{aligned}} \frac{\Scale[0.94]{C^{(m)}\big(\mathbf{L}(t),\mathbf{x}(t),N(t)\big)}}{\Scale[0.94]{C^*\big(\mathbf{L}(t),\mathbf{x}(t),N(t)\big)}},\label{PoAm}
\end{align}
which is greater than $1$. We next prove the PoA lower bound.

\begin{theorem}\label{thm:PoAm}
In the parallel traffic network with $M$ stochastic paths, as compared to the minimum social cost in (\ref{C*}), the myopic policy in (\ref{Cm}) results in: 
\begin{align}
    \Scale[0.93]{\text{PoA}^{(m)}}\geq \frac{\Scale[0.93]{2(1-\rho^k)}}{\Scale[0.93]{2-\rho-\rho^k}},\label{PoA>2}
\end{align}
where $k=1+\log_{\alpha_H}\left(M\frac{\big(\ell_0-\frac{N_{min}}{M}-V(\frac{N_{min}}{M})\big)(\alpha_H-1)}{\alpha_H N_{max}}+1\right)$. The lower bound in (\ref{PoA>2}) is larger than $2$ as $\rho\rightarrow 1$ and $k\rightarrow \infty$ (by setting $V(\frac{N_{min}}{M})\ll\ell_0$ and $\ell_0\gg \alpha_H \frac{N_{max}}{M}$).
\end{theorem}
Inspired by Proposition \ref{Prop:explore}, the worst case may happen when the myopic policy under-explores with strong hazard belief $x_i(t)$. {Initially, we set $\Scale[0.94]{\mathbb{E}[\alpha_i(0)|x_i(0)]=1}$ and expected travel costs $\Scale[0.94]{c_i(0)=c_0(N_{max})}$ for any stochastic path $i$. Then myopic users always choose safe path 0 to make $\Scale[0.94]{n_i^{(m)}(t)=0}$ in (\ref{nm(t)}).} However, the socially optimal policy frequently asks $\Scale[0.94]{n_i^*(t)>0}$ of new user arrivals to explore path $i$ to learn $\Scale[0.94]{\alpha_L=0}$, which greatly reduces future travel latency on this path. After that, all users will keep exploiting path $i$ with $\alpha_i(t)=\alpha_L$, and the expected travel cost on this path may gradually increase to $\Scale[0.94]{c_i(\frac{N_{min}}{M})=c_0(0)}$ again after at least $k$ time slots. As $\Scale[0.94]{k\rightarrow \infty}$ (by setting small observation error $\Scale[0.94]{V(\frac{N_{min}}{M})\ll\ell_0}$ and large latency $\Scale[0.94]{\ell_0\gg\alpha_H \frac{N_{max}}{M}}$) and $\rho \rightarrow 1$, we have $\Scale[0.94]{\text{PoA}^{(m)}\geq 2}$.

Note that the $\text{PoA}^{(m)}$ lower bound in (\ref{PoA>2}) decreases with observation error $\Scale[0.94]{V(\cdot)}$, aligning with Lemma \ref{lemma:n*-nm} that $n_i^{(m)}(t)$ approaches to $\Scale[0.94]{n_i^*(t)}$ as $\Scale[0.94]{V(\cdot)}$ enlarges. Besides, if $\Scale[0.94]{M\rightarrow \infty}$, then $k$ approaches infinity. In this case, the PoA lower bound in (\ref{PoA>2}) becomes the minimum $2$. as more stochastic paths mitigate congestion from users' exploration.
Since users' myopic routing can at least double the social cost with $\Scale[0.94]{\text{PoA}^{(m)}\geq 2}$, we are well motivated to design an efficient mechanism to regulate.

\section{CHAR Mechanism with Learning Convergence}\label{section5}
In this section, we first show that existing information-hiding and deterministic-recommendation mechanisms do not work with even $\text{PoA}=\infty$. 
Accordingly, we propose our new CHAR mechanism and prove its minimum possible $\text{PoA}<\frac{5}{4}$.

Before that, we first demonstrate in the following that, the socially optimal policy, instead of the myopic policy, ensures correct long-run learning convergence of hazard belief $x_i(t)$.
\begin{proposition}\label{Prop:optimal_stationary}
Under optimal exploration number $n_i^*(t)$, once $\Scale[0.94]{V(N_{min})<\ell_0}$, as $\Scale[0.94]{t\rightarrow \infty}$, $\Scale[0.94]{x_i(t)}$ is guaranteed to converge to its real steady state $\Scale[0.94]{\Bar{x}}$ in (\ref{bar_alpha}), while the myopic $n^{(m)}_i(t)$ cannot. 
\end{proposition}
If $\Scale[0.94]{V(N_{min})\geq \ell_0}$, the huge observation error discourages the system to explore and converge to steady state $\Bar{x}$ in any way. Therefore, we only consider $\Scale[0.94]{V(N_{min})< \ell_0}$.
Under the myopic policy, if $\Scale[0.94]{\mathbb{E}[\ell_i(t)|x_i(t)]> \ell_0}$ with $\Scale[0.94]{\mathbb{E}[\alpha_i(t)|x_i(t)]\geq 1}$ for any stochastic path $i$, all users will never explore stochastic paths, and $\Scale[0.94]{x_i(t)}$ may deviate a lot from $\Scale[0.94]{\Bar{x}}$. 
While the socially optimal policy is willing to frequently explore stochastic path~$i$ to learn a low-hazard state $\alpha_L$ for followers. With long-term frequent exploration, hazard belief $x_i(t)$ can finally converge to its real steady state $\Bar{x}$. The convergence to $\Bar{x}$ also helps the system to decide optimal routing and reduce social costs.

\subsection{Benchmark Informational Mechanisms Comparison}\label{section:benchmark}

In practice, informational mechanisms are non-monetary and easier to implement as compared to pricing (e.g., \cite{li2022online,Ferguson2022effective}). Then we analyze the efficiency of two widely used informational mechanisms: information-hiding \cite{tavafoghi2017informational,wang2020efficient,farhadi2022dynamic} and deterministic-recommendation \cite{li2019recommending,wu2019learning,li2023congestion}. Note that we also tried prior Bayesian persuasion \cite{das2017reducing,kamenica2019bayesian,mansour2022bayesian}, which focus on one-shot games and cannot optimize long-run information learning.

Based on Proposition \ref{Prop:optimal_stationary}, if the platform hides all the information, i.e., $\mathbf{L}(t),\mathbf{x}(t)$ and $N(t)$, as \cite{tavafoghi2017informational,wang2020efficient,farhadi2022dynamic}, users can only estimate that each stochastic path has reached its steady state $\Scale[0.94]{\Bar{x}\sim \mathbb{P}(\Bar{x})}$ in (\ref{bar_alpha}) and $\Scale[0.94]{\mathbb{E}[N(t)]=N}$ for any time $t$. As in \cite{tavafoghi2017informational}, expected exploration number $n_i^{\emptyset}\in\{0,1,\cdots,N(t)\}$ of stochastic path $i$ under information-hiding becomes constant:
\begin{align}
    \Scale[0.94]{n_i^{\emptyset}=}\min
    \left\{\frac{\Scale[0.94]{N(t)}}{\Scale[0.94]{M}},\frac{\Scale[0.94]{N+c_0(0)-\mathbb{E}_{\Bar{x}\sim \mathbb{P}(\Bar{x})}[c_i(0)|\Bar{x}]}}{\Scale[0.94]{M+1}}\right\}, \label{n_empty}
\end{align}
where $\Scale[0.94]{c_0(\cdot)}$ and $\Scale[0.94]{c_i(\cdot)}$ are defined in (\ref{cm_1}) and (\ref{cm_2}), respectively.
We next prove the infinite PoA caused by the hiding mechanism.
\begin{lemma}\label{lemma:information_hiding}
If the platform hides all the information, i.e., $\mathbf{L}(t),\mathbf{x}(t)$ and $N(t)$, from users, the constant policy $n^{\emptyset}_i$ in (\ref{n_empty}) makes users either under- or over-explore stochastic path~$i$ as compared to the social optimum, leading to $\text{PoA}=\infty$.
\end{lemma}
If $\Scale[0.94]{\mathbb{E}_{\Bar{x}\sim \mathbb{P}(\Bar{x})}[c_i(\frac{N_{max}}{M})|\Bar{x}]\leq c_0(0)}$, all the $\Scale[0.94]{N(t)}$ users without information will choose stochastic paths with less expected cost. However, the actual $\Scale[0.94]{\alpha_i(t)}$ can be $\Scale[0.94]{\alpha_H}$ and $\Scale[0.94]{\mathbb{E}[\ell_i(t)|x_i(t)]\gg \ell_0}$, then users’ maximum over-exploration makes PoA arbitrarily large. Similarly, if $\Scale[0.94]{\mathbb{E}_{\Bar{x}\sim \mathbb{P}(\Bar{x})}[c_i(0)|\Bar{x}]> c_0(N_{max})}$, all users will choose path 0, leading to the maximum under-exploration.

Next, we consider existing deterministic recommendation mechanisms (\!\!\cite{li2019recommending,wu2019learning,li2023congestion}), which privately provide state-dependent deterministic recommendations to proper users while hiding other information, to prove the caused infinite PoA. 
\begin{lemma}\label{lemma:deterministic_recommend}
The deterministic-recommendation mechanism makes $\text{PoA}=\infty$  in our system.
\end{lemma}
When a user receives recommendation $\Scale[0.94]{\pi(t)=i}$, he only infers $\Scale[0.94]{n_i(t)\geq 1}$ on path $i$. However, if expected user number $\Scale[0.94]{N\gg 1}$, this information is insufficient to alter his posterior distribution of $\Scale[0.94]{\Bar{x}}$ from $\Scale[0.94]{\mathbb{P}(\Bar{x})}$. Consequently, the caused exploration number still approaches $\Scale[0.94]{n^{\emptyset}_i}$ in (\ref{n_empty}), leading to $\Scale[0.94]{\text{PoA}=\infty}$. 

\subsection{New CHAR Mechanism Design and Analysis}

Inspired by Section \ref{section:benchmark}, the previous two mechanisms do not work in improving our system performance. Thus, we will properly combine them for a new mechanism design. To approach optimal policy $n_i^*(t)$ as much as possible, we dynamically select a number $\Scale[0.94]{N^{\emptyset}(t)}$ of users to follow hiding policy $n_i^{\emptyset}$ in (\ref{n_empty}), while providing state-dependent probabilistic recommendations to the remaining $\Scale[0.94]{N(t)-N^{\emptyset}(t)}$ users.

Finally, we are ready to propose our CHAR mechanism, which contains two steps per time slot $t$. Under CHAR, let $\Scale[0.94]{n^{(\text{CHAR})}_i(t)\in\{0,\cdots,N(t)\}}$ represent the exploration number of path $i$ at $t$. Similar to (\ref{Cm}), denote $\Scale[0.94]{C^{(\text{CHAR})}(\mathbf{L}(t),\mathbf{x}(t),N(t))}$ to be the long-term social cost under $\mathbf{n}^{(\text{CHAR})}(t)$.
\begin{definition}[\textbf{CHAR mechanism}] \label{def:CHAR}
At any $t$, in the first step, the platform randomly divides $\Scale[0.94]{N(t)}$ users into the hiding-group with $\Scale[0.94]{N^{\emptyset}(t)}$ users and the recommendation-group with $\Scale[0.94]{N(t)-N^{\emptyset}(t)}$ users, where $\Scale[0.94]{N^{\emptyset}(t)}$ is the optimal solution to 
\begin{align}
    \min_{N^{\emptyset}(t)\in\{0,\cdots,N(t)\}} &\Scale[0.92]{C^{(\text{CHAR})}(\mathbf{L}(t),\mathbf{x}(t),N(t))},\label{Cc}\\
    \text{s.t.} \quad\quad\ \ &\Scale[0.92]{n_i^{(\text{CHAR})}(t)=N^{\emptyset}(t)}\cdot\frac{\Scale[0.92]{n_i^{\emptyset}}}{\Scale[0.92]{N(t)}}+\label{n^*:N_H}\\ &\quad\quad\quad\ \ \Scale[0.92]{(N(t)-N^{\emptyset}(t))\mathbf{Pr}\big(\pi(t)=i|x_i(t)\big)}\notag,
\end{align}
where
\begin{align}
    \Scale[0.92]{\mathbf{Pr}\big(\pi(t)=i|x_i(t)\big)=}\begin{cases}
        \Scale[0.92]{p_L},\ \text{ if }\Scale[0.92]{x_i(t)< x_{th}},\\
        \Scale[0.92]{p_H},\text{ if }\Scale[0.92]{x_i(t)\geq x_{th}},
    \end{cases}\label{Pr(pi=2)}
\end{align}
with $x_{th}$ is in Proposition \ref{Prop:explore} and any feasible $\Scale[0.94]{p_L}$ and $\Scale[0.94]{p_H}$ to satisfy $p_L\mathbb{P}(x_{th})\geq p_H(1-\mathbb{P}(x_{th}))$.

In the second step, the platform performs as follow:
\begin{itemize}
    \item For the $\Scale[0.94]{N^{\emptyset}(t)}$ hiding-group users, the platform hides all the past information from them. 
    \item For the rest $\Scale[0.94]{N(t)-N^{\emptyset}(t)}$ users in the recommendation-only group, the platform randomly recommends a path $\pi(t)$ to each user arrival by the following distribution:
    \begin{align}
        \pi(t)=\begin{cases}
            i,&\text{with }\Scale[0.92]{\mathbf{Pr}(\pi(t)=i|x_i(t))} \text{ in (\ref{Pr(pi=2)})},\\
            0,&\text{with }\Scale[0.92]{1-\sum_{i=1}^M\mathbf{Pr}(\pi(t)=i|x_i(t))}.
        \end{cases}\label{recommend_pi}
    \end{align}
\end{itemize}
\end{definition}
According to Definition \ref{def:CHAR}, the $N^{\emptyset}(t)$ users in the hiding-group rely on hiding policy $n_i^{\emptyset}$ in (\ref{n_empty}) to make routing decisions. For the recommendation-only group, given the always feasible $p_L$ and $p_H$ to satisfy $p_L \mathbb{P}(x_{th})\geq p_H (1-\mathbb{P}(x_{th}))$, if a user receives a recommendation $\pi(t)=i$, he infers that the posterior hazard belief satisfies $x_i(t)<x_{th}$ on path $i$. Thus, this user will not deviate from the recommendation for a smaller expected travel cost. Similarly, if recommended $\pi(t)=0$, the user can infer that any stochastic path $i$ is more likely to have $x_i(t)\geq x_{th}$ and accordingly will choose safe path 0. 

Given Bayesian incentive compatibility for all users, according to (\ref{n^*:N_H}), there are respectively $\Scale[0.94]{N^{\emptyset}(t)\cdot n_i^{\emptyset}/N(t)}$ users in the hiding-group and $\Scale[0.94]{(N(t)-N^{\emptyset}(t))\cdot \mathbf{Pr}(\pi(t)=i|x_i(t))}$ users in the recommendation-only group traveling on any path~$i$. By optimizing $\Scale[0.92]{C^{(\text{CHAR})}(\mathbf{L}(t),\mathbf{x}(t),N(t))}$ in (\ref{Cc}) to derive $\Scale[0.94]{N^{\emptyset}(t)}$, our CHAR mechanism makes $n_i^{(\text{CHAR})}(t)$ approach optimum $n_i^*(t)$ on each stochastic path~$i$, which avoids myopic policy's zero-exploration in Theorem \ref{thm:PoAm} and benchmark mechanisms' maximum-exploration of bad traffic conditions in Lemma \ref{lemma:information_hiding}. Thus, our CHAR mechanism efficiently reduces $\Scale[0.94]{\text{PoA}^{(m)}\geq 2}$ caused by the myopic policy to the minimal possible value.
\begin{theorem}\label{thm:poa=1}
Our CHAR mechanism in Definition \ref{def:CHAR} ensures Bayesian incentive compatibility for all users and achieves 
\begin{align}
    \Scale[0.94]{\text{PoA}^{(\text{CHAR})}=1+\frac{1}{2(M+1)\big(1+\frac{M}{N}\cdot V\big(\frac{N(2M+1)}{2M(M+1)}\big)\big)}},\label{PoA_c}
\end{align}
which is always less than $\frac{5}{4}$ and cannot be further reduced by any informational mechanism.
\end{theorem}

Under the stationary distribution $\Scale[0.94]{\mathbb{P}(\Bar{x})}$, if the travel cost of path 0 with any flow satisfies $c_0(0)\geq c_i\big(\frac{N_{max}}{M}\big)$ for any stochastic path $i$, all users will opt for choosing stochastic paths. In this case, no informational mechanism can curb their maximum-exploration. Despite this, users under our CHAR mechanism only over-explore stochastic paths in good traffic conditions ($\alpha_L$), thereby attaining the minimal possible PoA. 

Our CHAR mechanism's PoA in (\ref{PoA_c}) increases with the expected user number $\Scale[0.94]{N}$ and decreases with path number $\Scale[0.92]{M}$. As $\Scale[0.92]{N\rightarrow \infty}$, $\Scale[0.92]{\text{PoA}^{(\text{CHAR})}}$ coverges to $1+\frac{1}{2(M+1)}$ with minimum $\frac{5}{4}$. While if $\Scale[0.92]{M\rightarrow\infty}$, $\Scale[0.92]{\text{PoA}^{(\text{CHAR})}}$ approaches the optimum $1$.

\begin{figure}[t]
    \centering
    \includegraphics[width=0.35\textwidth]{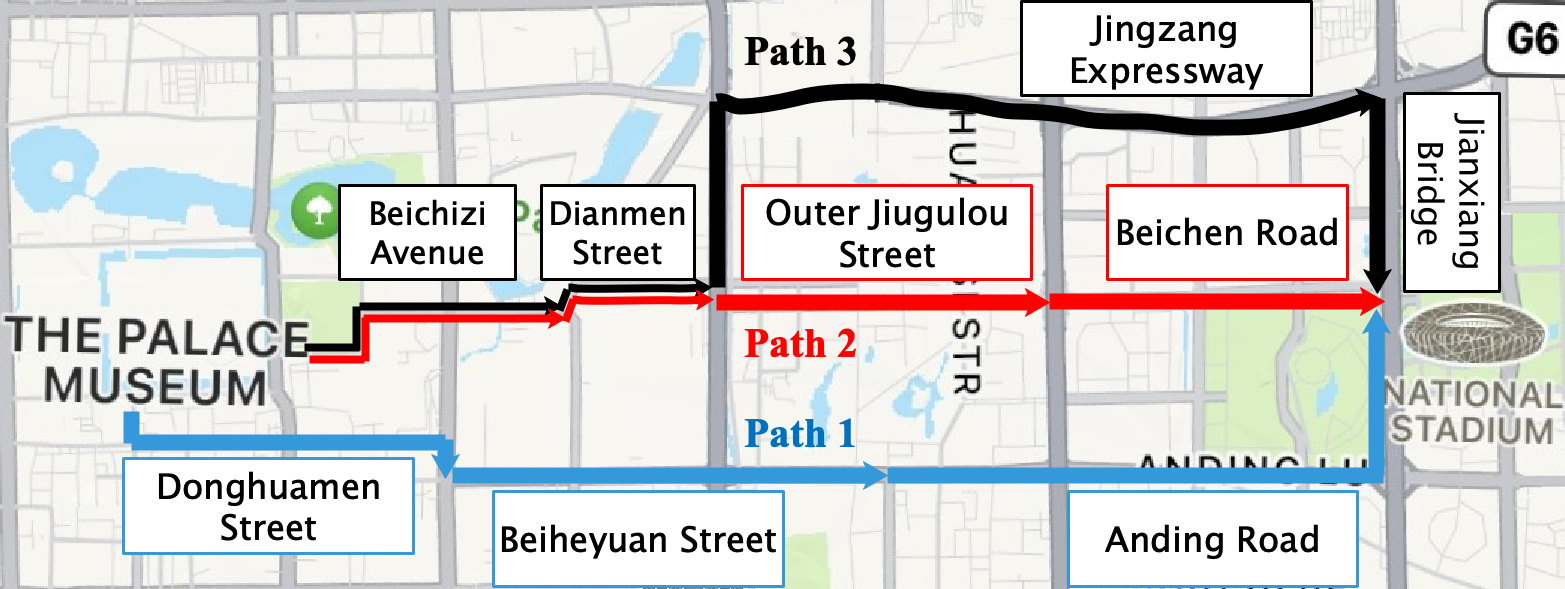}
    \captionsetup{font={footnotesize}}
    \caption{A popular hybrid road network consisting of three path choices from the Palace Museum to the National Stadium.}
    \label{fig:map}
\end{figure}

\section{Experiment Validation Using Real Datasets}\label{section6}
In this section, we experiment with real-world data to verify our CHAR's average performance versus the myopic policy (used by Waze) and the hiding mechanism (\!\!\cite{tavafoghi2017informational,wang2020efficient,farhadi2022dynamic}). To further practicalize our congestion model in (\ref{bar_alpha}), we sample peak hours' real-time traffic congestion data in Beijing, China on public holidays using BaiduMap dataset \cite{Baidumap}, and extend our parallel network in Fig.~\ref{fig:congestion_game} to the hybrid network in Fig. \ref{fig:map}.

\begin{figure}[t]
    \centering
    \includegraphics[width=0.3\textwidth]{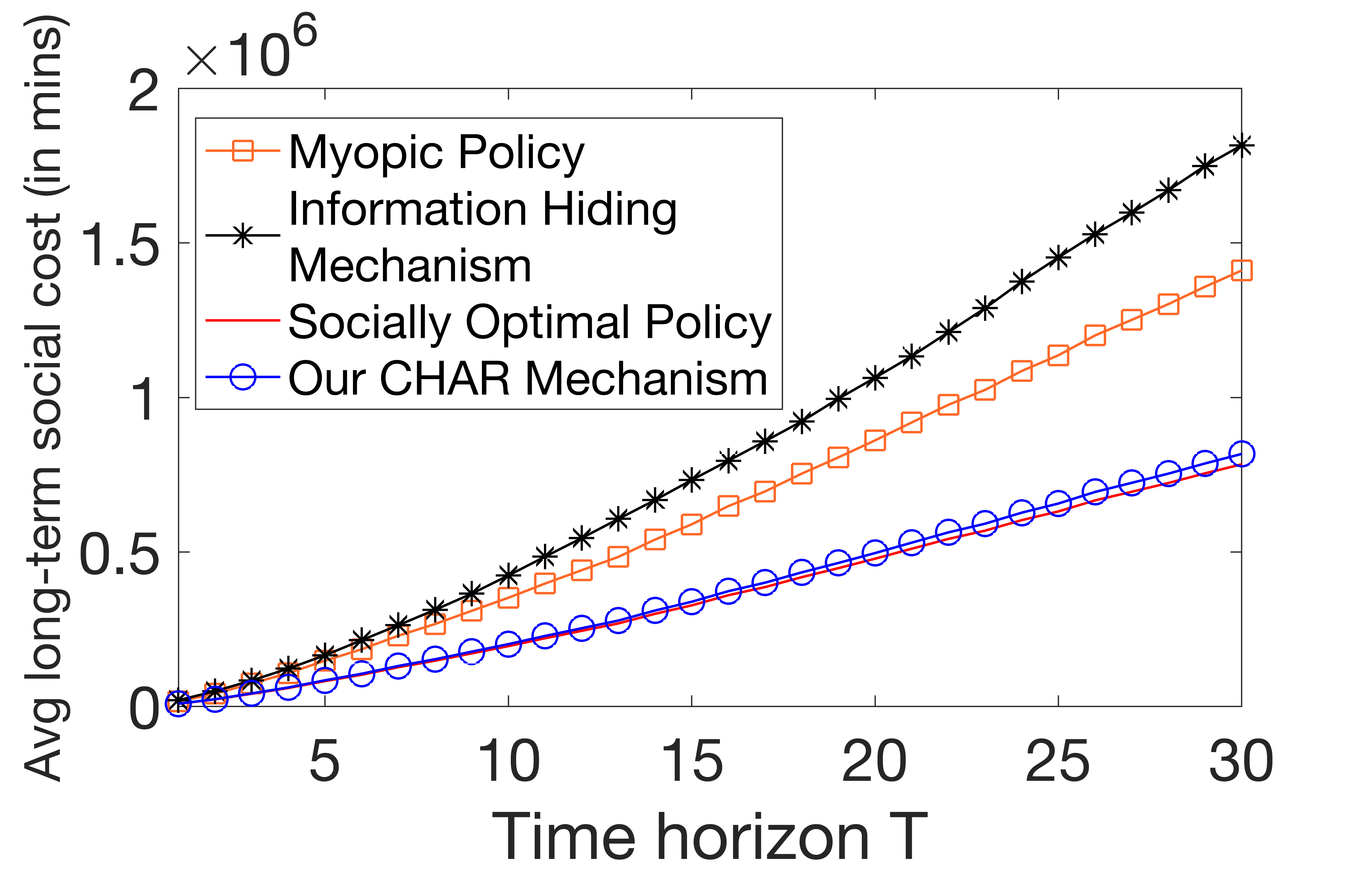}
    \captionsetup{font={footnotesize}}
    \caption{Average long-term costs (in minutes) under myopic, information hiding and socially optimal policies, and our CHAR mechanism versus $T$.}
    \label{fig:avg_cost}
\end{figure}

In Fig.~\ref{fig:map}, we validate that the traffic conditions of Donghuamen Street and Beiheyuan Street on path~1, Beichizi Avenue on both paths~2 and 3, and Jianxiang Bridge on path~3 can be well approximated as Markov chains with two discretized states (high and low) as in (\ref{bar_alpha}), while the other five roads tend to have deterministic conditions. Similar to \cite{eddy1998profile,chen2016predicting,wang2022framework}, we employ the hidden Markov model (HMM) approach to train the congestion model for the four stochastic road segments. 

We compare average long-term social costs under myopic, hiding, socially optimal policies, and our CHAR mechanism versus time horizon $T$. 
Fig.~\ref{fig:avg_cost} illustrates that our CHAR has less than $2\%$ efficiency loss from the social optimum for any time horizon $T$, while myopic and information-hiding policies cause around $75\%$ and $125\%$ efficiency losses, respectively.

\section*{Acknowledgment}

This work is supported by the Ministry of Education, Singapore, under its Academic Research Fund Tier 2 Grant with Award no. MOE-T2EP20121-0001. It is also supported by SUTD Kickstarter Initiative (SKI) Grant with no. SKI 2021\_04\_07 and the Joint SMU-SUTD Grant with no. 22-LKCSB-SMU-053.



\balance

\newpage

\appendix

\subsection{Proof of Lemma 1}
Given $M=1$, a myopic user will compare immediate travel costs of the two paths to choose the one that minimizes his own travel cost. Then there are three cases for the Nash equilibrium: 

If $c_1(0)\geq c_0(N(t))$, all the $N(t)$ users flock to safe path 0 with $n_1^{(m)}(t)=0$, which is the first case of (9). If $c_1(N(t))\leq c_0(0)$, all the $N(t)$ users flock to stochastic path~1 with $n_1^{(m)}(t)=N(t)$, which is the second case. Otherwise, the travel cost of each path is $c_0(N(t)-n_1^{(m)}(t))=c_1(n_1^{(m)}(t))$ in equilibrium. Solving this equation, we obtain the third case:
    \begin{align}
        {n_1^{(m)}(t)=}\frac{{N(t)}}{{2}}+\frac{{c_0(0)- c_1(0)}}{{2}}.\notag
    \end{align}

\subsection{Proof of Lemma 2}
We first prove the monotonicity of exploration numbers $n^{(m)}_i(t)$ and $n^*_i(t)$ with respect to hazard belief $x_i(t)$ and error cost $V(\cdot)$. Based on the results, we then prove the monotonicity of $n^{(m)}_i(t)-n^*_i(t)$. Here we take the basic case $M=1$ with one stochastic path as an example to analytically prove the monotonicity. For the other cases with $M\geq 2$, the monotonicity still holds due to the homogeneity of stochastic paths. 

\subsubsection{Monotonicity of Exploration Numbers}
We first discuss the monotonicity of $n_1^{(m)}(t)$ based on its definition in (10). 
By combining(10) with (8) and (9), we rewrite $n_1^{(m)}(t)$ to
\begin{align}\label{22}
    \frac{\Scale[0.92]{N(t)+\ell_0-\mathbb{E}[\ell_1(t)|x_1'(t-1)]-V(n_1^{(m)}(t-1))}}{2}.\tag{21}
\end{align}
It is obvious that $n_1^{(m)}(t)$ linearly decreases with $\mathbb{E}[\ell_1(t)|x_1'(t-1)]$ and $V(\cdot)$. Therefore, given expected latency $\mathbb{E}[\ell_1(t-1)|x'_1(t-2)]$ at the last time slot, if hazard belief $x_1(t)$ increases, the current $\mathbb{E}[\ell_1(t)|x_1'(t-1)=x_1(t)]$ also becomes longer, which consequently reduces exploration number $n_1^{(m)}(t)$.

Then we further prove the monotonicity of $n_1^*(t)$ by solving the long-term objective function under the socially optimal policy in (13). Here we focus on the general case with $n_1^*(t)\geq 1$ to calculate $n_1^*(t)$. As $n_1^*(t)$ is derived at the extreme point of the cost function $C^*\big(\mathbb{E}[\ell_1(t)|x_1'(t-1)],x_1(t),N(t)\big)$, we need to solve the first-order-derivative condition $\frac{\partial C^*\big(\mathbb{E}[\ell_1(t)|x_1'(t-1)],x_1(t),N(t)\big)}{\partial n_1^*(t)}=0$. Then we obtain
\begin{align}
    n_1^*(t)=&\frac{N(t)}{2}+\frac{\ell_0-\mathbb{E}[\ell_1(t)|x_1'(t-1)]-V(n_1^*(t-1))}{4}\notag\\-&\frac{\rho}{4}\frac{\partial \mathbb{E}[C^*(\mathbb{E}[\ell_1(t+1)|x_1'(t)],x_1(t+1),N)]}{\partial n_1^*(t)}.\label{23}\tag{22}
\end{align}
Next, we prove $n_1^*(t)$'s decreases with $V(n_1^*(t-1))$ by:
\begin{align*}
    \frac{\partial n^*(t)}{\partial V(n^*(t-1))}
    =&-\frac{1}{4}-\frac{\rho}{4}\frac{\partial\frac{\partial \mathbb{E}[C^*(\mathbb{E}[\ell_1(t+1)|x_1'(t)],x_1(t+1),N)]}{\partial n^*(t)}}{\partial V(n_1^*(t-1))},
\end{align*}
which is less than $0$ due to $\frac{\partial\frac{\partial \mathbb{E}[C^*(\mathbb{E}[\ell_1(t+1)|x_1'(t)],x_1(t+1),N)]}{\partial n^*(t)}}{\partial V(n_1^*(t-1))}=0$, as current observation error $V(n_1^*(t-1))$ has no effect on cost-to-go since the next time slot. Based on the above analysis, we obtain that $n_1^*(t)$ decreases with $V(\cdot)$. Then we can use the same method to prove $\frac{\partial n_1^*(t)}{\partial x_1(t)}\leq 0$ to show that $n_1^*(t)$ also decreases with hazard belief $x_1(t)$. 

\subsubsection{Monotonicity of $n^{(m)}_i(t)-n^*_i(t)$}
Based on the monotonicity of $n^{(m)}_i(t)$ and $n^*_i(t)$ above, we will prove $n_i^*(t)-n_i^{(m)}(t)$ increases with $x_i(t)$ by showing $\frac{\partial (n_i^*(t)-n_i^{(m)}(t))}{\partial x_i(t)}\geq 0$. If $M=1$, according to (\ref{22}) and (\ref{23}), we obtain
\begin{align}\label{24}
    &\frac{\partial (n_1^*(t)-n_1^{(m)}(t))}{\partial x_1(t)}\notag\\=&-\frac{1}{4}\frac{\mathbb{E}[\ell_1(t)|x_1'(t-1)]}{\partial x_1(t)}+\frac{1}{2}\frac{\mathbb{E}[\ell_1(t)|x_1'(t-1)]}{\partial x_1(t)}\notag\\&-\frac{\rho}{4}\frac{\partial^2 \mathbb{E}[C^*(\mathbb{E}[\ell_1(t+1)|x_1'(t)],x_1(t+1),N)]}{\partial n_1^*(t)\partial x_1(t)}\notag\\ =&\frac{1}{4}\frac{\mathbb{E}[\ell_1(t)|x_1'(t-1)]}{\partial x_1(t)}\notag\\ &-\frac{\rho}{4}\frac{\partial^2 \mathbb{E}[C^*(\mathbb{E}[\ell_1(t+1)|x_1'(t)],x_1(t+1),N)]}{\partial n_1^*(t)\partial x_1(t)}.\tag{23}
\end{align}
Based on the socially optimal policy in (13), we expand $\mathbb{E}[C^*(\mathbb{E}[\ell_1(t+1)|x_1'(t)],x_1(t+1),N)]$ since $t+1$ as:
\begin{align*}
    &\mathbb{E}[C^*(\mathbb{E}[\ell_1(t+1)|x_1'(t)],x_1(t+1),N)]\\=&\mathbf{Pr}(y_1(t)=1|n^*_1(t))C^*(\mathbb{E}[\ell_1(t+1)|x_1'(t),y_1(t)=1])\\&+\mathbf{Pr}(y_1(t)=0|n^*_1(t))C^*(\mathbb{E}[\ell_1(t+1)|x_1'(t),y_1(t)=0])
\end{align*}
Then we calculate $\frac{\partial^2 \mathbb{E}[C^*(\mathbb{E}[\ell_1(t+1)|x_1'(t)],x_1(t+1),N)]}{\partial n_1^*(t)\partial x_1(t)}$ in (\ref{24}) as:
\begin{align*}
     &\frac{\partial^2\mathbb{E}[C^*(\mathbb{E}[\ell_1(t+1)|x_1'(t)],x_1(t+1),N)]}{\partial n_1^*(t)\partial x_1(t)}\\=&\frac{\partial^2 \mathbf{Pr}(y_1(t)=1|n_1^*(t))}{\partial n_1^*(t)\partial x_1(t)}C^*(\mathbb{E}[\ell_1(t+1|x_1'(t),1)])\\ &+\frac{\partial C^*(\mathbb{E}[\ell_1(t+1|x_1'(t),1)])}{\partial x_1(t)}\frac{\partial \mathbf{Pr}(y_1(t)=1|n_1^*(t))}{\partial n_1^*(t)}\\ &+\frac{\partial^2 \mathbf{Pr}(y_1(t)=0|n_1^*(t))}{\partial n_1^*(t) \partial x_1(t)} C^*(\mathbb{E}[\ell_1(t+1|x_1'(t),0)])\\ &+\frac{\partial C^*(\mathbb{E}[\ell_1(t+1|x_1'(t),0)])}{\partial x_1(t)}\frac{\partial \mathbf{Pr}(y_1(t)=0|n_1^*(t))}{\partial n_1^*(t)}.
\end{align*}
Taking the above equation back to (\ref{24}), we obtain
\begin{align*}
    &\frac{\partial (n_1^*(t)-n_1^{(m)}(t))}{\partial x_1(t)}\\ \geq& \frac{1}{4}\frac{\mathbb{E}[\ell_1(t)|x_1'(t-1)]}{\partial x_1(t)}\\&-\frac{\rho}{4}\frac{\partial C^*(\mathbb{E}[\ell_1(t+1|x_1'(t),1)])}{\partial x_1(t)}\frac{\partial \mathbf{Pr}(y_1(t)=1|n_1^*(t))}{\partial n_1^*(t)}\\ &+\frac{\rho}{4}\frac{\partial C^*(\mathbb{E}[\ell_1(t+1|x_1'(t),0)])}{\partial x_1(t)}\frac{\partial \mathbf{Pr}(y_1(t)=1|n_1^*(t))}{\partial n_1^*(t)}\\ \geq& \frac{1}{4}\frac{\mathbb{E}[\ell_1(t)|x_1'(t-1)]}{\partial x_1(t)}\\&-\frac{\rho}{4}\frac{\partial C^*(\mathbb{E}[\ell_1(t+1|x_1'(t),1)])}{\partial x_1(t)}\frac{\partial \mathbf{Pr}(y_1(t)=1|n_1^*(t))}{\partial n_1^*(t)}\\ &+\frac{\rho}{4}\frac{\partial C^*(\mathbb{E}[\ell_1(t+1|x_1'(t),1)])}{\partial x_1(t)}\frac{\partial \mathbf{Pr}(y_1(t)=1|n_1^*(t))}{\partial n_1^*(t)}\\ =&\frac{1}{4}\frac{\mathbb{E}[\ell_1(t)|x_1'(t-1)]}{\partial x_1(t)}>0,
\end{align*}
where the first inequality is because of $\mathbf{Pr}(y_1(t)=0|n_1^*(t))=1-\mathbf{Pr}(y_1(t)=1|n_1^*(t))$, and the last inequality is due to that $\mathbb{E}[\ell_1(t)|x_1'(t-1)]$ increases with $x_1(t)$. In summary, the difference $n_1^*(t)-n_1^{(m)}(t)$ increases with $x_1(t)$. 
 
Next, we use the same method to prove that $(n_1^*(t)-n_1^{(m)}(t))$ decreases with observation error function $V(\cdot)$:
\begin{align*}
    &\frac{\partial (n_1^*(t)-n_1^{(m)}(t))}{\partial V(n_1^*(t-1))}\\=&\frac{1}{4}-\frac{\rho}{4}\frac{\partial^2 \mathbb{E}[C^*(\mathbb{E}[\ell_1(t+1)|x_1'(t)],x_1(t+1),N)]}{\partial n^*(t)\partial V(n^*(t-1))},
\end{align*}
which is larger than $0$ as $\frac{\partial^2 \mathbb{E}[C^*(\mathbb{E}[\ell_1(t+1)|x_1'(t)],x_1(t+1),N)]}{\partial n_1^*(t)\partial V(n_1^*(t-1))}=0$. This completes the proof.

\subsection{Proof of Proposition 1}
Given $\mathbb{E}[\ell_i(t-1)|x_i'(t-2)]$ on stochastic path $i$ at $t-1$, we first prove that there exists a high belief $x_i(t)=R$ to make the myopic policy under-explore path $i$ (with $n_i^*(t)-n_i^{(m)}(t)>0$). Then we prove that there exists another low belief $x_i(t)=r<R$ to make the myopic policy over-explore (with $n_i^*(t)-n_i^{(m)}(t)\leq 0$). Based on Lemma~2 that $n_i^*(t)-n_i^{(m)}(t)$ increases with $x_i(t)$, there exists an unique exploration threshold $x_{th}\in(r,R)$, which also increases with $V(\cdot)$. 

\subsubsection{Proof of Under-exploration}\label{under-explore}
Let hazard belief $x_i(t)=R$ satisfies $\mathbb{E}[\alpha_i(t)|R]=1$ for any stochastic path $i\in\{1,\cdots,M\}$. If $\ell_0+N(t)\leq\mathbb{E}[\ell_i(t)|R]+V(0)$, all the $N(t)$ users under the myopic policy will choose safe path~0. Then the caused long-term expected social cost is
\begin{align*}
    C^{(m)}(\mathbf{L}(t),\mathbf{x}(t),N(t))=\sum_{t=0}^{\infty}\rho N(\ell_0+N)=\frac{N(\ell_0+N)}{1-\rho}.
\end{align*}
However, the socially optimal policy may still recommend $n^*_i(t)$ users to explore a stochastic path $i$ to obtain 
\begin{align}\label{25}
    &C^*(\mathbf{L}(t),\mathbf{x}(t),N(t)) \notag\\ \leq &n_i^*(t)\big(\mathbb{E}[\ell_i(t)|R]+n_i^*(t)+V(0)\big)+(N(t)-n_i^*(t))\ell_0 \notag\\ &+\rho \mathbb{E}_{y_i(t)}[C^*(\mathbf{L}(t+1),\mathbf{x}(t+1),N)].\tag{24}
\end{align}

In (\ref{25}), there are two cases to update cost-to-go $\mathbb{E}_{y_i(t)}[C^*(\mathbf{L}(t+1),\mathbf{x}(t+1),N)]$ in future time slots.
If $y_i(t)=1$, for any future user arrivals at $\tau>t$, the socially optimal policy will recommend $n_i^*(\tau)=0$ users to explore stochastic path $i$ to avoid extra travel cost. Thus,
\begin{align*}
    \Scale[0.94]{C^*(\mathbf{L}(t+1),\mathbf{x}(t+1),N|y_i(t)=1)}\leq& \frac{N(\ell_0+N)}{1-\rho}\\=&\Scale[0.94]{C^{(m)}(\mathbf{L}(t),\mathbf{x}(t),N)}.
\end{align*}
While if $y_i(t)=0$, the expected travel latency for $t+1$ becomes $0$ due to low-hazard state $\alpha_L=0$. As the cost function increases with travel latency on path $i$, the cost-to-go satisfies
\begin{align*}
    C^*(\mathbf{L}(t+1),\mathbf{x}(t+1),N|y_i(t)=0)<&C^*(\mathbf{L}(t),\mathbf{x}(t),N)\\\leq &C^{(m)}\big(\mathbf{L}(t),\mathbf{x}(t),N\big).
\end{align*}
In summary, the cost-to-go under socially optimal policy is always smaller than that under the myopic policy.
If $\ell_0\gg 0$, $N(t)\ll \ell_0$ and $V(N(t))\ll\ell_0$, we further calculate (\ref{25}):
\begin{align*}
    &C^*(\mathbf{L}(t),\mathbf{x}(t),N(t))
    \\\leq &n_i^*(t)\big(\mathbb{E}[\ell_i(t)|R]+n_i^*(t)+V(0)\big)+(N(t)-n_i^*(t))\ell_0\\ &+\rho \mathbf{Pr}(y(t)=0)C^*(\mathbf{L}(t+1),\mathbf{x}(t+1),N|y_i(t)=0)\\ &+\rho \mathbf{Pr}(y(t)=1)C^*(\mathbf{L}(t+1),\mathbf{x}(t+1),N|y_i(t)=1)\\
    <&N(t)\ell_0+\rho C^{(m)}(\mathbf{L}(t),\mathbf{x}(t),N)
    =C^{(m)}(\mathbf{L}(t),\mathbf{x}(t),N(t)).
\end{align*}
Hence, if $x_i(t)\geq R$, current $N(t)$ users under-explore stochastic path $i$ under the myopic policy.

\subsubsection{Proof of Over-exploration}\label{over-explore}
Next, we prove that there exists a smaller belief $x_i(t)=r<R$ leading to the myopic policy's over-exploration on path $i$.
Based on (9), if $\ell_0>\mathbb{E}[\ell_i(t)|r]+N(t)+V(N(t-1))$, all the $N(t)$ myopic users choose to explore stochastic path $i$, leading to $n_i^*(t)\leq n_i^{(m)}(t)$ (over-exploration). However, we will still try to prove there exist parameters to strictly make $n_i^*(t)< n_i^{(m)}(t)$.

Given the above $\mathbf{L}(t),\mathbf{x}(t)$ and $N(t)$, we calculate:
\begin{align*}
    \Scale[0.93]{C^{(m)}(\mathbf{L}(t),\mathbf{x}(t),N(t))=}&\Scale[0.93]{N(t)(\mathbb{E}[\ell_i(t)|r]+N(t)+V(N(t-1))}\\&\Scale[0.93]{+ \rho \mathbb{E}_{y_i(t)}[C^{(m)}(\mathbf{L}(t+1),\mathbf{x}(t+1),N)]},
\end{align*}
where the cost-to-go since next time slot $t+1$ satisfies
\begin{align}
    &\mathbb{E}_{y_i(t)}[C^{(m)}(\mathbf{L}(t+1),\mathbf{x}(t+1),N)]\tag{25}\label{26}\\=&\Scale[0.95]{\mathbf{Pr}(y_i(t)=0|N(t)) C^{(m)}(\mathbf{L}(t+1),\mathbf{x}(t+1),N|y_i(t)=0)}\notag\\
    &\Scale[0.95]{+\mathbf{Pr}(y_i(t)=1|N(t)) C^{(m)}(\mathbf{L}(t+1),\mathbf{x}(t+1),N|y_i(t)=1)}.\notag
\end{align}

\begin{figure*}
	\begin{subequations}
		\begin{align}
		      \text{PoA}^{(m)}\tag{26}\label{27}\geq& \frac{\frac{1-\rho^k}{1-\rho}(\ell_0(t)+N(t))\cdot N(t)}{N(t)\cdot(\mathbb{E}[\ell_i(t)|x_i'(t-1)]+N(t)+V(N(t)))+\sum_{j=1}^k(\frac{(\rho \alpha_H)^{j-1}-1}{\rho\alpha_H-1}\rho \alpha_H+N(t)+V(N(t)))\cdot N(t)}\\
            \geq &\frac{\frac{1-\rho^k}{1-\rho}N(t)\ell_0(t)}{N(t)\ell_0(t)+\frac{\ell_0(t)}{2}N(t)\frac{\rho-\rho^k}{1-\rho}}
            =\lim_{k\rightarrow \infty}\frac{2(1-\rho^k)}{2-\rho-\rho^k}=2.\notag
		\end{align}
	\end{subequations}
	{\noindent} \rule[-10pt]{18cm}{0.05em}
\end{figure*}

We first assume that the optimal exploration number $n^*_i(t)< N(t)$. Then, if the group observation probabilities satisfy $q_L(N(t))>0, q_H(N(t))=1$ and $\alpha_L=0$, we obtain
\begin{align*}
    \mathbf{Pr}(y_i(t)=1|N(t))=&rq_H(N(t))+(1-r)q_L(N(t))\\=&r+(1-r)q_L(N(t))\\>&\mathbf{Pr}(y_i(t)=1|n_i^*(t)),
\end{align*}
due to the fact that $q_L(n_i(t))$ decreases with $n_i(t)$. Similarly, the posterior belief under $y_i(t)=1$ satisfies
\begin{align*}
    x'_i(t|y_i(t)=1,N(t))=&\frac{rq_H(N(t))}{rq_H(N(t))+(1-r)q_L(N(t))}\\ >&x_i'(t|y_i(t)=1,n_i^*(t)),
\end{align*}
as $q_H(n(t))$ increases with $n(t)$. 

Then the cost-to-go under the myopic policy in (\ref{26}) satisfies
\begin{align*}
    &\mathbb{E}_{y_i(t)}[C^{(m)}(\mathbf{L}(t+1),\mathbf{x}(t+1),N)]\\=&\Scale[0.95]{\mathbf{Pr}(y_i(t)=0|N(t)) C^{(m)}(\mathbf{L}(t+1),\mathbf{x}(t+1),N|y_i(t)=0)}\\ 
    &+\Scale[0.95]{\mathbf{Pr}(y_i(t)=1|N(t)) C^{(m)}(\mathbf{L}(t+1),\mathbf{x}(t+1),N|y_i(t)=1)}\\
    >&\Scale[0.95]{\mathbf{Pr}(y_i(t)=0|n^*_i(t)) C^{(m)}(\mathbf{L}(t+1),\mathbf{x}(t+1),N|y_i(t)=0)}\\
    &\Scale[0.95]{+\mathbf{Pr}(y_i(t)=1|n^*_i(t)) C^{(m)}(\mathbf{L}(t+1),\mathbf{x}(t+1),N|y_i(t)=1)}\\
    =&\mathbb{E}_{y_i(t)}[C^*(\mathbf{L}(t+1),\mathbf{x}(t+1),N)].
\end{align*}
As the cost-to-go $\mathbb{E}_{y_i(t)}[C^{(m)}(\mathbf{L}(t+1),\mathbf{x}(t+1),N)]$ under $n^{(m)}_i(t)=N(t)$ is greater than $\mathbb{E}_{y_i(t)}[C^*(\mathbf{L}(t+1),\mathbf{x}(t+1),N)]$ under $n^*_i(t)<N(t)$, current $N(t)$ users under the myopic policy over-explore stochastic path $i$ if $x_i(t)\leq r<R$.

In summary, we have proved $n_i^*(t)> n_i^{(m)}(t)$ for $x_i(t)\geq R$ in Appendix \ref{under-explore} and $n_i^*(t)\leq n_i^{(m)}(t)$ for $x_i(t)\leq r$ in Appendix \ref{over-explore}. Based on Lemma 2 that $n_i^*(t)-n_i^{(m)}(t)$ increases with hazard belief $x_2(t)$, there must exist unique exploration threshold $x_{th}\in (r,R)$, which increases with $V(\cdot)$.

\subsection{Proof of Theorem 1}
We prove this theorem by analyzing the worst-case scenario with the myopic policy's zero-exploration. However, the socially optimal policy recommends some users explore path $i$ to find possible $\alpha_L$ and reduce travel costs for future users. 

Initially, we set $\ell_0=\mathbb{E}[\ell_i(t)|x_i(t)]+V(N(t))$, $\alpha_L=0$ and $\alpha_H>1$ with $\mathbb{E}[\alpha_i(t)|x_i(t)]=1$, such that myopic users will never explore any stochastic path $i$ to avoid long travel latency there. As there is no information update and $\mathbb{E}[\alpha_i(t)|x_i(t)]=1$, the expected travel latency on stochastic path $i$ remains unchanged. Then the caused long-term expected social cost is
\begin{align*}
    C^{(m)}(\mathbf{L}(t),\mathbf{x}(t),N(t))=\frac{N(\ell_0+N)}{1-\rho}.
\end{align*}

For the socially optimal policy, it lets $\frac{N(t)}{M}$ users explore each stochastic path $i$ to derive immediate social cost:
\begin{align*}
    c_i(n_i^*(t))=\frac{N(t)}{M}(\mathbb{E}[\ell_i(t)|x_i'(t-1)]+\frac{N(t)}{M}+V(0)).
\end{align*}

Suppose that the system has been running for a long time $t$ before the current time slot. Then the probability of the actual travel latency on stochastic path $i$ being reduced to $\mathbb{E}[\ell_i(t)]=0$ by low hazard state $\alpha_L=0$ is $\mathbf{Pr}(\mathbb{E}[\ell_i(t)]=0)=1-x_i(t)^t\rightarrow 0$.
Accordingly, it is almost sure for current $N(t)$ users to observe $\alpha_i(t)=\alpha_L$ on each path $i$. 
After that, the travel cost $c_i(\frac{N_{min}}{M})$ of path $i$ gradually increases to $c_0(0)$ again at after $k$ time slots. In the worst case, all the $N(t)=\frac{N_{max}}{M}$ users always observe $y_i(t)=1$ to make $\alpha(t)=\alpha_H$. Based on the linear correlation function of $\mathbb{E}[\ell_i(t+1)|x_i'(t)]$, the travel latency for $(t+k)$-th time slot satisfies
\begin{align*}
    \mathbb{E}[\ell_i(t+k)]\leq \sum_{j=1}^k \alpha_H^j \frac{N_{max}}{M}=\frac{\alpha_H^{k-1}-1}{\alpha_H-1}\alpha_H \frac{N_{max}}{M}.
\end{align*}
Based on this inequality, we solve $c_i(\frac{N_{min}}{M})=c_0(0)$ to obtain
\begin{align*}
    k\geq 1+\log_{\alpha_H}\left(M\frac{(\ell_1-\frac{N_{min}}{M}-V(\frac{N_{min}}{M})(\alpha_H-1)}{\alpha_H N_{max}}+1\right).
\end{align*}
It means that users' travel costs on the two paths become the same again after $k$ slots. Based on our analysis above, we calculate PoA in (\ref{27}), where the second inequality is due to the convexity in $k$ of the second term in the denominator.

\subsection{Proof of Proposition 2}
First, we prove that the myopic policy cannot ensure correct convergence by analyzing the same worst-case scenario as Theorem 1. Then we prove the socially optimum's convergence.

At the beginning of time $t$, if $\mathbb{E}[\alpha_i(t)|x_i'(t-1)]\geq 1$ and $\mathbb{E}[\ell_i|x_i'(t-1)]\geq \ell_0-V(n_i(t-1))$, selfish myopic users will never explore stochastic path $i$. In this case, $x_i(t)$ remains unchanged as $x_i'(t-1)$ and will never converge to $\Bar{x}$.

For the social optimum, we will prove that if $x_i(t)>\Bar{x}$, its updated $x_i(t+1)$ at $t+1$ is expected to decrease, i.e., $x_i(t+1)<x_i(t)$. While if $x_i(t)<\Bar{x}$, $x_i(t+1)$ is expected to be greater than $x_i(t)$. Based on the two cases, we obtain that $x_i(t)$ will finally converge to the real steady state $\Bar{x}$ under consequent explorations of stochastic paths.

If the actual hazard belief $x_i(t)>\Bar{x}$, then users' expected probability of observing a hazard is
\begin{align*}
    \Scale[0.93]{\mathbf{Pr}(y_i(t)=1|n(t),x_i(t))=}&\Scale[0.93]{(1-x_i(t))q_L(n_i(t))+x_i(t)q_H(n_i(t))}\\
    >& \Scale[0.93]{(1-\Bar{x})q_L(n_i(t))+\Bar{x}q_H(n_i(t))}\\
    =&\Scale[0.93]{\mathbf{Pr}(y_i(t)=1|n_i(t),\Bar{x})},
\end{align*}
based on the fact that $q_H(n_i(t))>q_L(n_i(t))$ and $x_i(t)>\Bar{x}$. It means the actual probability $\mathbf{Pr}(y_i(t)=1|n_i(t),\Bar{x})$ for current $n_i(t)$ users to observe a hazard ($y_i(t)=1$) under $\Bar{x}$ is lower than the expected probability $\mathbf{Pr}(y_i(t)=1|n_i(t),x_i(t))$. Similarly, we obtain $\mathbf{Pr}(y_i(t)=0|n_i(t),x_i(t)) <\mathbf{Pr}(y_i(t)=0|n_i(t),\Bar{x})$,
i.e., the actual probability to observe $y_i(t)=0$ under $\Bar{x}$ is greater than the expected probability under $x_i(t)$.

Based on the above analysis, if there are users traveling on stochastic path $i$ and sharing their observation summary $y_i(t)$, the actually updated belief at the next time slot $t+1$ satisfies
\begin{align*}
    &x_i(t+1)\\=&\mathbf{Pr}(y_i(t)=1|\Bar{x})x'_i(t|1)+\mathbf{Pr}(y_i(t)=0|\Bar{x})x'_i(t|0)\\
    <& \mathbf{Pr}(y_i(t)=1|x_i(t))x'_i(t|1)+\mathbf{Pr}(y_i(t)=0|x_i(t))x'_i(t|0)\\ =& \mathbb{E}_{y_i(t)}[x_i(t+1)|x_i(t)] = x_i(t),
\end{align*}
which means that the hazard belief $x_i(t)$ will actually decreases to $x_i(t+1)$ if $x_i(t)>\Bar{x}$. Similarly, we can prove that the actually updated belief $x_i(t+1)> x_i(t)$ if $x_i(t)<\Bar{x}$. 

In summary, under the socially optimal policy with users' consequent exploration and learning on stochastic path $i\in\{1,\cdots,M\}$, $x_i(t)$ can finally converge to $\Bar{x}$ as $t\rightarrow \infty$.

\subsection{Proof of Lemma 3}\label{Proof_G}

Under the information hiding mechanism, users can only infer that there are $\mathbb{E}[N(t)]=N$ users arriving currently. 

Initially, let $\ell_0\geq \mathbb{E}_{\Bar{x}\sim \mathbb{P}(\Bar{x})}[\Bar{\ell_i}|\Bar{x}]+V(0)+\frac{N_{max}}{M}$, such that $n_i^{\emptyset}=N(t)$ in (15), as $c_0(0)-\mathbb{E}_{\Bar{x}\sim \mathbb{P}(\Bar{x})}[c_i(0)|\Bar{x}]\geq \frac{N_{max}}{M}$. At the same time, we set actual $\mathbb{E}[\ell_i(0)|x_i(0)]\gg \ell_0$ and $\mathbb{E}[\alpha_i(0)|x_i(0)]>1$. In this case, the social optimum will recommend all the $N(t)$ users travel safe path 0 to reduce the latency on path $i$. The caused PoA under over-exploration is:
\begin{align*}
    \text{PoA}\geq \frac{\sum_{j=0}^\infty \rho^j N(\ell_i(0)+\frac{N}{M}+V(\frac{N}{M}))}{\sum_{j=0}^\infty \rho^j N\ell_0} =\infty.
\end{align*}

While if $c_0(N_{max})-\mathbb{E}_{\Bar{x}\sim \mathbb{P}(\Bar{x})}[c_i(0)|\Bar{x}]\leq -N_{max}$ and the initial actual travel latency on stochastic path $i$ satisfies $\mathbb{E}[\ell_i(0)|x_i(0)]\ll \ell_0$, then the information-hiding mechanism leads to under-exploration of stochastic path $i$.

\subsection{Proof of Lemma 4}

We prove that the recommendation-only mechanism makes PoA infinite by showing that, if $N\gg 1$, users will still follow $n_i^{\emptyset}$ under the information hiding policy in (15). 

Under the recommendation-only mechanism, if a user is recommended to choose stochastic path $\pi(t)=i$, his posterior distribution of the long-run expected belief $\Bar{x}$ changes from $\mathbb{P}(\Bar{x})$ to $\mathbb{P}'(\Bar{x}|n_i^*(t)\geq 1)$. Then he estimates the expected travel latency on stochastic path $i$ as $\mathbb{E}_{\Bar{x}\sim \mathbb{P}'(\Bar{x}|n_i^*(t)\geq 1)} [c_i(1)|\Bar{x}]\approx 
    \mathbb{E}_{\Bar{x}\sim \mathbb{P}(\Bar{x})} [c_i(1)|\Bar{x}],$
given $N\gg 1$. In consequence, this user will not follow this recommendation $\pi(t)=i$ if $\mathbb{E}_{\Bar{x}\sim \mathbb{P}(\Bar{x})} [c_i(1)|\Bar{x}]>c_0(N-1)$, even with all the other $N-1$ users choosing path 0. 
Similarly, if a user is recommended to choose safe path $\pi(t)=0$, his posterior distribution becomes $\mathbb{P}'(\Bar{x}|n_i^*(t)\geq 0)$, which always equals $\mathbb{P}(\Bar{x})$ without any other information. Therefore, each user still follows $n_i^{\emptyset}$ in (15) to make his path decision, leading to $\text{PoA}=\infty$ as in Lemma 3.

\subsection{Proof of Theorem 2}
We first prove that each user in the recommendation-only group is incentive compatible to follow recommendation $\pi(t)$, and there always exists $p_L$ and $p_H$ to satisfy the condition $p_L\mathbb{P}(x_{th})\geq p_H(1-\mathbb{P}(x_{th}))$. Then we prove that by dynamically solving (18) to decide number $N^{\emptyset}(t)$, our CHAR realizes the PoA in (20). Finally, we show that this PoA in (20) is the minimum achievable and cannot be reduced by any other informational mechanism. 

\subsubsection{Incentive Compatibility for all Users}
If a user of the recommendation-only group is recommended $\pi(t)=i$, he estimates that there will be $\Bar{n}_i^{(\text{CHAR})}$ users on path~$i$ under CHAR. Then his posterior probability of $x_i(t)<x_{th}$ becomes:
\begin{align*}
    &\mathbf{Pr}(x_i(t)<x_{th}|\pi(t)=i)\\=& \frac{\mathbf{Pr}(\pi(t)=i,x_i(t)<x_{th})}{\mathbf{Pr}(\pi(t)=i)}
    \\=&\frac{\mathbf{Pr}(\pi(t)=i|x_i(t)<x_{th})\mathbf{Pr}(x_i(t)<x_{th})}{\mathbf{Pr}(\pi(t)=i)}\\=&\frac{\mathbb{P}(x_{th})p_L}{\mathbb{P}(x_{th})p_{L}+(1-\mathbb{P}(x_{th}))p_H},
\end{align*}
where the last equality is derived by replacing $\mathbf{Pr}(\pi(t)=i)$ with $\mathbf{Pr}(\pi(t)=i|x_i(t)<x_{th})\mathbf{Pr}(x_i(t)<x_{th})+\mathbf{Pr}(\pi(t)=i|x_i(t)\geq x_{th})\mathbf{Pr}(x_i(t)\geq x_{th})$ in the second equality, according to the law of total probability.
Similarly, his estimated posterior probability $\mathbf{Pr}(x_i(t)\geq x_{th}|\pi(t)=i)$ of $x_i(t)\geq x_{th}$ is $\frac{(1-\mathbb{P}(x_{th}))p_H}{\mathbb{P}(x_{th})p_{L}+(1-\mathbb{P}(x_{th}))p_H}$.

Given $p_L\mathbb{P}(x_{th})\geq p_H(1-\mathbb{P}(x_{th}))$, the above two probabilities satisfy $\mathbf{Pr}(x_i(t)<x_{th}|\pi(t)=i)>\mathbf{Pr}(x_i(t)\geq x_{th}|\pi(t)=i)$. Then this user further compares the expected cost of choosing stochastic path $i$ and the cost of safe path 0:
\begin{align}\label{29}
    &\Scale[0.94]{\mathbb{E}_{x_i(t)}[c_i(\Bar{n}_i^{(\text{CHAR})})]-\mathbb{E}_{x_i(t)}[c_0(\Bar{n}_0^{(\text{CHAR})})]}\tag{28}\\
    =&\int_0^1(c_i(\Bar{n}_i^{(\text{CHAR})})- c_0(\Bar{n}_0^{(\text{CHAR})}))d\mathbb{P}'(\Bar{x}|\pi(t)=i)\notag\\
    =&\int_0^{x_{th}}(c_i(\Bar{n}_i^{(\text{CHAR})})- c_0(\Bar{n}_0^{(\text{CHAR})}))d\mathbb{P}'(\Bar{x}|\pi(t)=i)\notag\\&+\int_{x_{th}}^1(c_i(\Bar{n}_i^{(\text{CHAR})})- c_0(\Bar{n}_0^{(\text{CHAR})}))d\mathbb{P}'(\Bar{x}|\pi(t)=i).\notag
\end{align}
Let $\lambda_1(\Bar{x})<0$ and $\lambda_2(\Bar{x})>0$ denote the average values of two integration above. Since the cost functions $c_0(\cdot)$ and $c_i(\cdot)$ are concave with respect to $x_i(t)$, the two average values satisfy $|\lambda_1(\Bar{x})|> \lambda_2(\Bar{x})$. Taking them back to (\ref{29}), we obtain
\begin{align*}
    (\ref{29}) =&\Scale[0.94]{\lambda_1(\Bar{x})\mathbb{P}'(x_{th}|\pi(t)=i)+\lambda_2(\Bar{x})(1-\mathbb{P}'(x_{th}|\pi(t)=i))}\\=& \lambda_1(\Bar{x})\mathbf{Pr}(x_i(t)<x_{th}|i)+\lambda_2(\Bar{x})\mathbf{Pr}(x_i(t)\geq x_{th}|i)\\ \leq & \lambda_1(\Bar{x})+\lambda_2(\Bar{x}) \leq 0,
\end{align*}
where the first inequality is because of the posterior probabilities $\mathbf{Pr}(x_i(t)<x_{th}|\pi(t)=i)>\mathbf{Pr}(x_i(t)\geq x_{th}|\pi(t)=i)$ under the condition $p_L\mathbb{P}(x_{th})\geq p_H(1-\mathbb{P}(x_{th}))$, and the last inequality is due to $\lambda_1(\Bar{x})+\lambda_2(\Bar{x})<0$ given $x_i(t)\leq x_{th}$. Therefore, users given recommendation $\pi(t)=i$ believe that the expected cost on path $i$ with $\Bar{n}^{(\text{CHAR})}_i$ users is less than the cost of switching to path 0 with $\Bar{n}^{(\text{CHAR})}_0$ users there. As a result, they will not deviate from the recommendation $\pi(t)=i$. Under our assumption that $\mathbb{P}(\Bar{x})$ is mild, users may travel on any stochastic path $i$ under $\Bar{x}\sim \mathbb{P}(\Bar{x})$. Thus, there always exists $p_L,p_H\in (0,1)$ to satisfy $p_L\mathbb{P}(x_{th})\geq p_H(1-\mathbb{P}(x_{th}))$.

Similarly, if a user is recommended to choose safe path 0, his expected travel cost on path 0 is less than the cost on any stochastic path $i$, due to $\mathbf{Pr}(x_i(t)\geq x_{th}|\pi(t)=0)>\mathbf{Pr}(x_i(t)< x_{th}|\pi(t)=0)$. In summary, all users in the recommendation-only group are incentive compatible to follow the recommendations, and the expected users choosing path $i$ in this group is $(N(t)-N^{\emptyset}(t))\mathbf{Pr}(\pi(t)=i|x_i(t))$.

Regarding users in the hiding-group, they have to use their prior distribution $\mathbb{P}(\Bar{x})$ to estimate the expected costs on the two paths and make decisions as $n^{\emptyset}_i$ in (15). Therefore, the expected number of users choosing path $i$ in the hiding group is $\frac{n_i^{\emptyset}}{N(t)} N^{\emptyset}(t)$, and they are also incentive-compatible. 

\subsubsection{$\text{PoA}^{(\text{CHAR})}$ in (21)}
Recall Proposition 1 that myopic users will under-explore stochastic path $i$ if $x_i(t)\geq x_{th}$ and over-explore if $x_i(t)<x_{th}$. If $x_i(t)\geq x_{th}$ with a bad condition on stochastic path $i$, the hiding policy satisfies $n_i^{\emptyset}>n_i^*(t)$ and the recommendation probability satisfies $N(t)p_L<n_i^*(t)$, such that our CHAR can always change $n_i^{(\text{CHAR})}(t)$ to be the optimal $n_i^*(t)$ by dynamically changing the user number $N^{\emptyset}(t)$ of the hiding-group, according to the exploration number $n_i^{(\text{CHAR})}(t)$ in (17). While if $x_i(t)<x_{th}$ with a good traffic condition on path $i$, both the hiding policy and the recommendation probability satisfy $n_i^{\emptyset}>n_i^*(t)$ and $p_H N(t)>n_i^*(t)$, such that $n_i^{(\text{CHAR})}>n_i^*(t)$. In consequence, the worst-case scenario under our CHAR is users' maximum over-exploration. 

In the worst-case scenario, the expected exploration number $n_i^{(\text{CHAR})}(t)$ under our CHAR mechanism is 
\begin{align}
    n_i^{(\text{CHAR})}(t)=\frac{N}{M}=\frac{N}{M+1}+\frac{c_0(0)-c_i(0)}{M+1}, \tag{29}\label{30}
\end{align}
for any $t$. The caused immediate social cost per time slot is
\begin{align}
    c(\mathbf{n}^{(\text{CHAR})}(t))&=\sum_{i=1}^M \frac{N}{M} c_i\left(\frac{N}{M}\right)=N c_i\left(\frac{N}{M}\right). \tag{30}\label{31}
\end{align}
While the socially optimal policy will let some users choose path 0 to avoid the congestion on each stochastic path $i\in\{1,\cdots,M\}$. Suppose that the system has been running for a long time, such that the socially optimal policy $n^*_i(t)$ has reached its steady state. Therefore, solving $\frac{\partial c(n^*_i(t))}{\partial n_i^*(t)}=0$, we obtain the optimal expected exploration number
\begin{align}
    n^*_i(t)=\frac{N}{M+1}+\frac{c_0(0)-c_i(0)}{2(M+1)}\tag{31}\label{32}
\end{align}
for any stochastic path $i$ and $n^*_0(t)=\frac{N}{2(M+1)}$ for safe path 0.
Then we calculate the corresponding immediate social cost as:
\begin{align}
    c(\mathbf{n}^*(t))&=n^*_0(t)c_0(n^*_0(t))+\sum_{i=1}^M n^*_i(t) c_i(n^*_i(t))\tag{32}\label{33}\\
    &=\frac{N}{2(M+1)}(c_0(0)+c_i(0))+\frac{M}{M+1}Nc_i(0).\notag
\end{align}
As the immediate cost under our CHAR mechanism in (\ref{31}) and that under the socially optimal policy in (\ref{33}) remain unchanged for any time slot $t$, we calculate the expected PoA as
\begin{align*}
    \text{PoA}^{(\text{CHAR})}&=\frac{c(\mathbf{n}^{(\text{CHAR})}(t))}{c(\mathbf{n}^*(t))}\\ &=1+\frac{c_0(0)-c_i(0)}{c_0(0)+c_i(0)+2Mc_i(0)}\\
    &=1+\frac{1}{2(M+1)(1+\frac{M}{N}V(\frac{N(2M+1)}{2M(M+1)}))},
\end{align*}
by inputting $c_0(0)=\frac{N}{M}+c_i(0)$ in (\ref{30}) and $n_i^*(t)$ in (\ref{32}).

Under the stationary distribution $\mathbb{P}(\Bar{x})$, if the travel cost of path 0 with any flow satisfies $c_0(0)\geq c_i(\frac{N_{max}}{M})$ for any stochastic path $i$, all users will opt for choosing stochastic paths, i.e., $n^{\emptyset}_i$ in (15) satisfies $n^{\emptyset}_i=\frac{N}{M}=n^{(m)}_i(t)$. Then any informational mechanism cannot change their expected travel cost of stochastic path $i$, and thus cannot curb their over-exploration. Therefore, $\text{PoA}^{(\text{CHAR})}$ in (21) is the minimum achievable PoA by any informational mechanism.

\subsection{Experiment Details}
In this experiment, we mined the datasets from Baidu Map using its provided API (BaiduMap 2023). Please use Google Translate to translate it into English and follow the interface documentation to access data using API.

As depicted in Figure 2, we consider a popular hybrid road network from the Palace Museum to the National Stadium on public holidays, as many visitors randomly leave the Palace Museum and travel to the National Stadium. We use the traffic flow data from Baidu Map to calculate the mean arrival car number is $N=121$ every $5$ minutes at the gateway of the Palace Museum, with a standard deviation of $12.33$. 

To train a practical congestion model with travel latency, we mine and analyze many data about the real-time traffic status values of the nine roads, which can dynamically change every $5$ minutes in the peak-hour periods. The traffic status values vary from $1$ to $4$ to tell the real-time congestion levels. We validate from the dataset that the traffic conditions of Donghuamen Street and Beiheyuan Street on path 1, Beichizi Avenue on both paths 2 and 3, and Jianxiang Bridge on path 3 can be well approximated as Markov chains with two discretized states (high and low traffic states) as in (2), while the other five roads tend to have deterministic/safe conditions. Similar to (Eddy 1998; Chen, Wen and Geng 2016), we employ the hidden Markov model (HMM) approach to train the transition probability matrices for the four stochastic road segments using $432$ statuses, which suffice to achieve high accuracy. The details of data processing and training are provided below.
\begin{itemize}
    \item Using the raw data extracted from Baidu Map, we assigned manual labels to each traffic status value. If the value corresponds to a good traffic condition (i.e., $1$ or $2$), it is labeled as $1$. If the value corresponds to a bad traffic condition (i.e., $3$ or $4$), it is labeled as $2$. The labeled data is then stored in an immediate data file.
    \item In MATLAB, we begin by reading the immediate data file to store raw data and their labels into two lists for each road, such as $\text{Donghuamen}_{\text{seq}}$ and $\text{Donghuamen}_\text{state}$ for Donghuamen Street. Subsequently, we employ the $\textbf{hmmestimate}(\cdot)$ function to conduct training and derive the transition probability matrices for all stochastic paths.
    \item Based on the transition probability matrices, we calculate the steady state $\Bar{x}$ of each road as follow:
    \begin{align*}
        &\text{Donghuamen\_x}=0.3883,  \text{ Beihe\_x}\quad\ =0.1064,\\ &\text{Beichizi\_x}\quad\ \ \ \ =0.1915, \text{ Jianxiang\_x}=0.9362.
    \end{align*}
    Then, such steady-state $\Bar{x}$ of each road is used in our later simulations as the congestion model. 
    \item Finally, based on the travel latency datasets, we calculate the long-term average correlation coefficients $\alpha_H\approx 1.3$ and $\alpha_L\approx 0.3$, by assuming the linear latency function.
\end{itemize}

Besides the congestion model above, we set discount factor $\rho=0.98$ for the long-term social costs and initial hazard belief as $0.5,0.2,0.3,$ and $0.8$ for Donghuamen Street, Beiheyuan Street, Beichizi Avenue, and Jianxiang Bridge, respectively.

\end{document}